\documentclass[aps,pra,superscriptaddress,twocolumn]{revtex4}
\usepackage{eurosym}
\usepackage{amsmath}
\usepackage{amssymb}
\usepackage{bm}
\usepackage{epsfig}
\usepackage{graphicx}
\usepackage{color}
\usepackage{inputenc}

\newcommand{\wt}[1]{\widetilde{#1}}

\begin{document}

\title{Spin-orbit coupled soliton in a random potential}
\author{Sh. Mardonov}
\affiliation{Department of Physical Chemistry, The University of the Basque Country
UPV/EHU, 48080 Bilbao, Spain}
\affiliation{The Samarkand Agriculture Institute, 140103 Samarkand, Uzbekistan}
\affiliation{The Samarkand State University, 140104 Samarkand, Uzbekistan}
\author{{V. V. Konotop}}
\affiliation{Centro de F\'{i}sica Te\'orica e Computacional, Faculdade de
	Ci\^encias and Departamento de F\'{i}sica, Faculdade de Ci\^encias, Universidade
	de Lisboa, Campo Grande, Ed. C8, Lisboa 1749-016, Portugal}
\author{{B. A. Malomed}}
\affiliation{{{Dept. of Physical Electronics, School of Electrical Engineering
and Center for Light-Matter Interaction, 
Tel Aviv University, Tel Aviv 69978, Israel}}}
\author{M. Modugno}
\affiliation{{Dept. of Theoretical Physics and History of Science, University of the
Basque Country UPV/EHU, 48080 Bilbao, Spain}}
\affiliation{IKERBASQUE Basque Foundation for Science, Bilbao, Spain}
\author{E. Ya. Sherman}
\affiliation{Department of Physical Chemistry, The University of the Basque Country
UPV/EHU, 48080 Bilbao, Spain}
\affiliation{IKERBASQUE Basque Foundation for Science, Bilbao, Spain}

\date{\today}

\begin{abstract}
We investigate theoretically the dynamics of a spin-orbit coupled soliton formed by a self-interacting 
Bose-Einstein condensate immersed in a random potential, in the presence of an artificial magnetic field. 
We find that due to the anomalous spin-dependent velocity, the synthetic Zeeman coupling 
can play a critical role in the soliton dynamics by
causing its localization or delocalization, depending on the coupling strength 
and on {the parameters of the random potential}. The observed effects of the 
Zeeman coupling qualitatively depend on the type of self-interaction in the condensate
since the spin state and the self-interaction energy of the condensate are mutually related if the 
invariance of the latter with respect to the spin rotation is lifted.
\end{abstract}

\maketitle

\section{Introduction} 

The ability to emulate spin-orbit coupling (SOC) and the Zeeman interaction 
in Bose-Einstein condensates (BECs) \cite{Spielman2009,Splielman2011,Zhai2012,Spielman2013,Zhai2015} 
has raised a great interest in
the interplay of nonlinear phenomena and spin dynamics of these systems. 
These effects include the 
creation of solitons {\cite{Kevrekidis,KaKoAb13,KaKoZe14,Wen2016,Chiquillo2018}}, 
vortices \cite{vortices0,vortices1,vortices2,LoKaKo14,vortices3,Busch},
localized spinful structures \cite{BenLi,HPu,Sakaguchi2016,Vasic2016,Romania}, 
enhanced localization \cite{Qu2017}, {bound states in continuum~\cite{BIC}}, 
and  collapsing solutions \cite{collapse,Yu2017a,Yu2017b}. This research 
greatly extends the understanding of solitons in other systems, such as 
nonlinear photonic lattices \cite{pertsch2004,kartashov2005,kartashov2008,kartashov2009,kartashov2011,naether2013}. 
In addition, the studies of disorder potentials which can be produced experimentally, 
have demonstrated a strong qualitative interplay between nonlinearity and quantum 
localization \cite{larcher2009,Flach2,Aleiner2010}. 

In this paper we address the motion of 
a bright soliton in a BEC with attractive interaction, the dynamics of which 
can be strongly affected by the disorder and the 
SOC, even in the semiclassical regime (as considered in the present paper), where quantum 
effects are not sufficiently strong to induce Anderson localization \cite{fort2005,modugno2006,Mardonov2015}. 
These effects can be experimentally observable to show
how the soliton propagation in a random potential can be affected by the
SOC and the condensate self-interaction.

\section{BEC-soliton in a random potential: model and main parameters.} 

We consider a quasi one-dimensional BEC \cite{Modugno2018}  with SOC forming a soliton 
due to the internal self-attraction and affected by a 
synthetic Zeeman field and {by a spin-diagonal disordered potential} \cite{Larcher2012}. The
two-component spinor wave function ${\bm\psi} (\mathbf{x})\equiv \left[ \psi _{1
}(\mathbf{x}),\psi _{2 }(\mathbf{x})\right]^{\rm T}$, where $\mathbf{x}\equiv(x,t),$ characterizing a pseudo-spin
$1/2$, is normalized to {unity}, and obtained as a solution of the time-dependent Gross-Pitaevskii equation 
\begin{equation}
i\hbar \partial _{t}{\bm\psi} =\left[\frac{\hbar^{2}\hat{k}^{2}}{2M}+\alpha {\sigma }_{z}%
\hat{k}+\frac{\Delta }{2}{\sigma }_{x}+U (x)+H^{\rm int}\right]{\bm\psi}, 
\label{Hamilton}
\end{equation}%
with the self-interaction term: 
$H^{\rm int}_{\lambda\lambda}=g\left|\psi _{\lambda}\right|^{2}+
\widetilde{g}\left|\psi_{\lambda^{\prime}}\right|^{2}$, and $H^{\rm int}_{\lambda\lambda^{\prime}}=0,$
where $\lambda,\lambda^{\prime}=1,2$, and $\lambda\ne\lambda^{\prime}.$
Here $M$ is the particle mass, $\hat{k}=-i\partial/\partial x$, 
$\alpha $ is the SOC constant, ${\sigma}_{z}$ and ${\sigma }_{x}$ 
are Pauli matrices, $\Delta $ is the Zeeman splitting, $U(x)$ is the random potential,
and $g$ and $\widetilde{g}$ are the interaction constants including the total number of atoms 
in the condensate. Hereafter we use the units with $M=\hbar\equiv\,1.$  
 
The intra-component coupling $g$ is assumed to be negative, $g<0$, and equal for the 
two components. 
The inter-component coupling $\widetilde{g}$, will be considered for two limiting cases.
First, for $g=\widetilde{g}$ 
the system self-interaction energy is invariant with respect to  the global spin rotations \cite{Manakov,Tokatly}. 
In this case and in the absence of $U(x)$ and spin-related interactions with $\Delta=\alpha=0,$ 
the ground state is given by: 
\begin{eqnarray}
\label{eq:soliton}
{\bm \psi}_{\rm gr}=\frac{\sqrt{-g}}{2\cosh\left[ 2(x-x_{0})/g\right] }\left[\begin{array}{c}
\cos(\theta/2)e^{i\phi}
\\
\sin(\theta/2)
\end{array}
\right],
\end{eqnarray}
where $x_0$ is a position of the soliton center and angles $\theta$ and $\phi$ characterize the pseudospin direction.
Second, we consider the case of a vanishing cross-spin coupling with $\widetilde{g}=0,$ where this invariance is lifted.  

The potential $U(x)$ is produced by
$N\gg 1$ ``impurities'' of the amplitude $U_{0}$ with uncorrelated random  positions $x_{j}$ 
as:
\begin{equation}
U (x)=U_{0}\sum_{j=1}^{N}s_{j}f\left(x-x_{j}\right). 
\label{randomrealization}
\end{equation}%
Here $s_{j}=\pm 1$ is a random function of $j$ with $\sum_{j=1}^{N}s_{j}=0$, resulting in the spatially averaged 
$\langle U(x)\rangle =0$. The mean 
linear density of impurities is given by $\bar{n}=N/L$, where $L$ is the sample length.  
The shape of a single impurity is given by $f\left(z\right) =\exp \left( -z^{2}/\xi ^{2}\right)$ 
with constant $\xi\ll L.$

In order to describe the dynamics of a soliton (or, more generally, a localized wavepacket) in the 
random potential we explore the integral quantities $\mathcal{O}(t)$ associated with each observable $\hat{\mathcal{O}}$ and defined by 
%The time-dependent expectation values of any quantity $\mathcal{O}$ are determined by ${\bm\psi}(\mathbf{x})$ as:
\begin{equation}
\mathcal{O}(t)=\int_{-\infty}^{\infty}{\bm\psi}^{\dagger}(\mathbf{x})
\hat{\mathcal{O}}
{\bm\psi}(\mathbf{x})dx.
\label{expectation}
\end{equation}
In particular, defining the total soliton momentum $k(t)$ and the force  $F(t)$, for which 
$\hat{\mathcal{O}}$ in (\ref{expectation}) is substituted by $\hat{k}$ and by $\hat{F}\equiv-dU(x)/dx$, respectively,
and using Eq. (\ref{Hamilton}), it is straightforward to verify the Ehrenfest-like relation
\begin{equation}
\frac{d k(t)}{dt}=F(t). 
\label{force}
\end{equation}

\begin{figure}[tb]
\begin{center}
\includegraphics[width=0.40\textwidth]{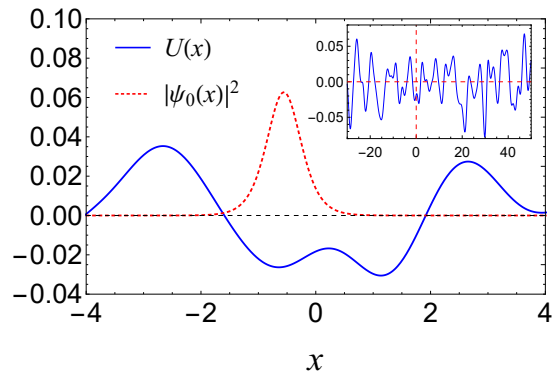} 
\end{center}
\caption{Density of the initial state (in arbitrary units) for $g=\widetilde{g}=-5$, for
a realization of disorder with $U_{0}=0.01$, $\bar{n}=10,$ and $\xi =1$, shown in the inset.}
\label{Fig:disorder10}
\end{figure}

As a reference point for the {following discussion}
we consider as the initial state an eigenstate of the 
Hamiltonian (\ref{Hamilton})  with $\alpha =\Delta =0$, of the form ${\bm\psi}_{\rm in}(x)=\psi
_{0}(x)\left[ 1,0\right]^{\rm T}$, i.e. $\psi _{0}(x)$ is a stationary solitonic solution in the
$U(x)-$potential. To produce this solution, we start with the state (\ref{eq:soliton}) with $\theta=\phi=0$
and adiabatically switch on $U(x)$, in order to project the initial soliton into a 
stationary state at equilibrium with the random potential. 
Figure \ref{Fig:disorder10} shows a realization of $U(x)$ 
and the density of the soliton prepared with this protocol. 
This soliton is localized near a potential minimum and subsequent 
dynamics is induced by switching on the SOC and the Zeeman field. 

We explore the spin state with the density matrix ${\bm \rho}(t):$   
\begin{equation}
{\bm\rho}(t)=\int {\bm\psi}(\mathbf{x}){\bm\psi}^{\dagger}(\mathbf{x})dx.
\label{denmat}
\end{equation}%
The rescaled purity $P(t)=2{\rm tr}{\bm \rho}^{2}(t)-1$ is the square of the spin length,
$P(t)=\sum_{i}(\sigma_{i}(t))^{2},$ with the spin components 
${\sigma}_{i}(t)={\mathrm{tr}\left({\sigma}_{i}{\bm \rho(t)}\right)},$ which can also be obtained with
Eq. \eqref{expectation}. 

In order to characterize the evolution of the system we {consider} the 
center of mass position $X(t)$ defined with $\hat{\mathcal{O}}=x$ 
in Eq. (\ref{expectation}). Then, the velocity of the wavepacket, described by Eq. (\ref{Hamilton}),  
defined as $v(t)=dX(t)/dt$, is given by the relation
\begin{equation}
{v}(t) = {k}(t) +\alpha \sigma _{z}(t). 
\label{spinvelocity}
\end{equation}%
Notice that this formula, which includes the anomalous term $\alpha\sigma_{z}(t),$ well-known in the linear theory	
\cite{Adams,Stepanov,Armaitis2017}, remains valid for nonlinear model (\ref{Hamilton}). 
According to Eq. (\ref{force}),
the soliton momentum evolves due to the random potential. 
Correspondingly, this  contributes to the spin precession due to the SOC.  

Since the Hamiltonian (\ref{Hamilton}) depends on many parameters, we consider only 
the case of a ``narrow'' soliton, having the width much less than $\xi$, and 
assuming that it is stable against
the collapse due to the presence of the {transverse} degrees of freedom \cite{Perez1997}. 
For a smooth random potential, the radiation from such a soliton is negligible, its dynamics is 
close to adiabatic, and the derivative $dX(t)/dt$ truly characterizes the soliton velocity. 
{Remarkably, for the} chosen parameters, the disorder potential has almost no effect on the shape of the
soliton. Here, the effective potential $V(t)$, computed with  Eq. (\ref{expectation}) 
for $\hat{\mathcal{O}}=U(x)$ is very close to $U(X(t)),$   
and for a weak random potential, where $\alpha\gg\,k(t),$ one estimates
$k(t)\approx\alpha\left[U(X(0))-U(X(t))\right],$ as follows from Eqs. (\ref{force}) and (\ref{spinvelocity}). 
For the numerical analysis we consider 
a single typical realization of disorder in Fig. \ref{Fig:disorder10}, and
use $\xi$ as the unit of length. With the accepted units $M\equiv\hbar\equiv 1,$
the units of energy and time become $1/\xi^{2}$ and $\xi^{2}$, respectively. 

\section{Motion of soliton}

\subsection{Localization by Zeeman field.}

We begin with the symmetric $g=\widetilde{g}$ case. 
{For $\Delta=0$ and $\alpha >0$}, the spin component 
{$\sigma_{z}(t)$ is conserved, i.e. $\sigma_{z}(t)\equiv 1$ for the initial condition ${\bm\psi}_{\rm in}(x)$ } and the soliton 
will start to displace to the right, owing to the spin dependent
velocity (\ref{spinvelocity}). For a small $\alpha,$ the
soliton undergoes harmonic oscillations in the vicinity of the initial position 
{because ${\bm\psi}_{\rm in} (x)$ is centered in the  local minimum of the potential}. 
However, for $\alpha$ larger than a critical value, as discussed below 
the soliton {moves} over long distances until it encounters a sufficiently strong peak 
{of the potential}, that can stop it and reverse its motion resulting in essentially nonlinear oscillations. 
\begin{figure}[tb]
\begin{center}
\hspace{-0.5cm}\includegraphics*[width=0.36\textwidth]{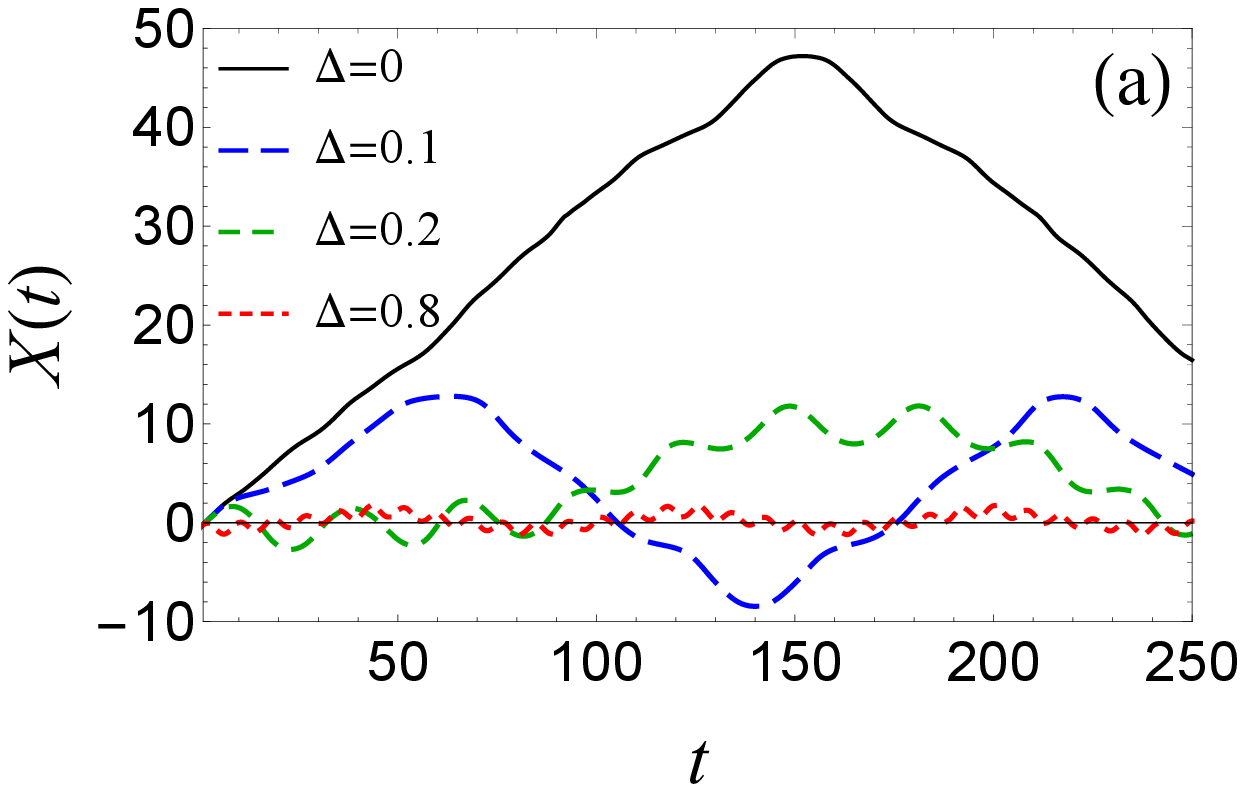} \\
\includegraphics*[width=0.4\textwidth]{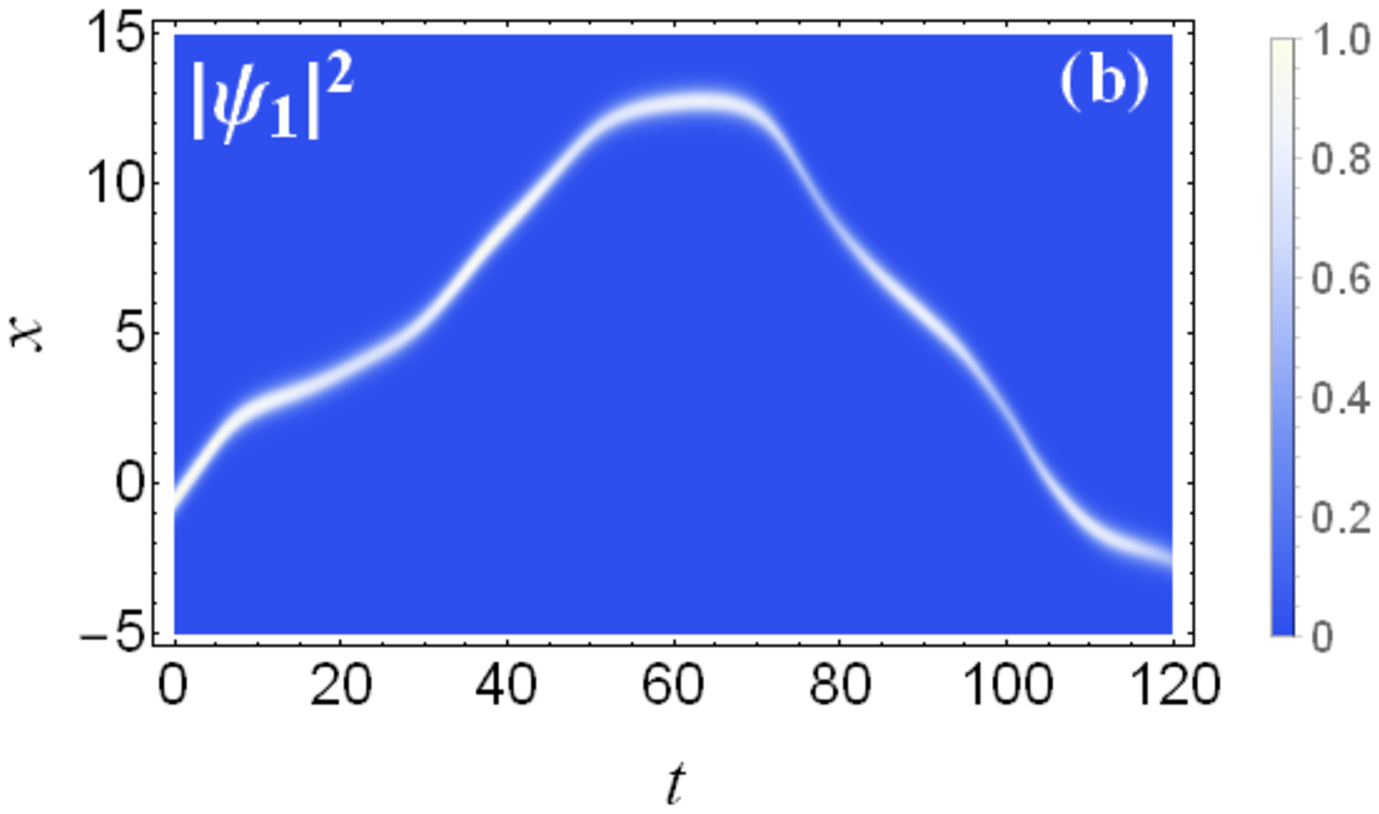} \\
\includegraphics*[width=0.4\textwidth]{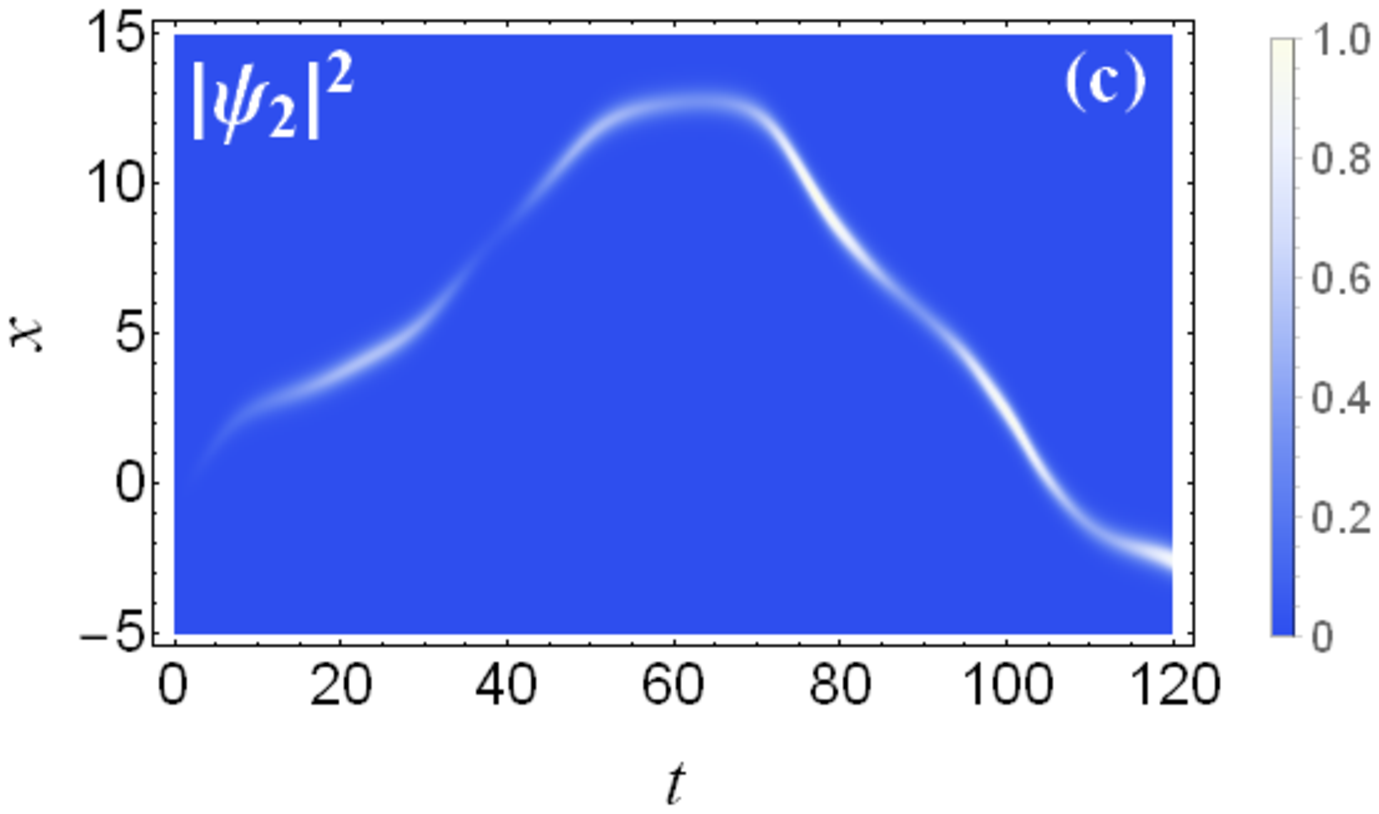}
\end{center}
\caption{(a) Position of the soliton center of mass as a function of time for different
$\Delta$ (marked in the plot), $g=\widetilde{g}=-5,$ and $\alpha =0.4$. Panel (a) shows that
for $\Delta =0$ the soliton travels a long distance, whereas switching on the
Zeeman field eventually traps it. (b, c) Density plots of the {two} spinor
components in the $(t,x)-$plane for $\Delta=0.1$. 
}
\label{Fig:x_local}
\end{figure}

Figure \ref{Fig:x_local} describes the dynamics of the soliton for $\alpha =0.4$
at different values of $\Delta$. 
The plot shows that, whereas for $\Delta=0$ the soliton moves a long distance 
through the disordered potential until it is reflected by a large fluctuation
of $U(x)$, the presence of a Zeeman field inhibits the propagation, 
and eventually traps it, for sufficiently large values of $\Delta$ ($=0.8$ in this plot).

In order to better understand this
effect of the Zeeman field, we consider {Eq. (\ref{spinvelocity}) for the velocity, along with} 
the evolution of the {spin components}. To describe the spin evolution we assume the adiabatic 
approximation for the soliton evolution  (conserving the shape of equal densities of both spin states):
${\bm\psi}_{\rm ad}({\mathbf x})=\psi_{0}(x-X(t))\exp(ik(t)x)\chi(t),$
where $\chi(t)\equiv\left[\cos(\theta(t)/2)e^{i\phi(t)},\sin(\theta(t)/2)\right]^{\rm T}$ 
describes corresponding evolution of the spin state. Here ``fast'' degree of freedom
corresponds to the shape of $\psi_{0}(x-X(t))$, 
%with the energy corresponding to $H^{\rm[is]}$ in Eq. \eqref{Hamilton}
while ``slow'', lower energy degrees of freedom are  described by $k(t)$ and $\chi(t)-$dependencies. 
%Thus, eliminating the fast motion
In order to define the adiabatic evolution of the spinor $\chi(t)$ we perform 
spatial ``averaging'' by multiplying (\ref{Hamilton}) by $\psi_{0}(x-X(t))$ and integrating over $x.$ This gives us an
effective Hamiltonian $H_{s}(t)=(\bm{\Omega}(t){\bm\sigma})/2$ for the spin motion in a synthetic
random Zeeman field as  ${\bm\Omega}=\left(\Delta ,0,2\alpha{k(t)}\right)$, 
corresponds to the rotation with the rate $\Omega=(4\alpha^{2}k^{2}(t)+\Delta^{2})^{1/2}$
around the randomly time-dependent axis ${\mathbf n}={\bm\Omega}/\Omega$.  

\begin{figure}[h]
\begin{center}
\includegraphics[width=0.40\textwidth]{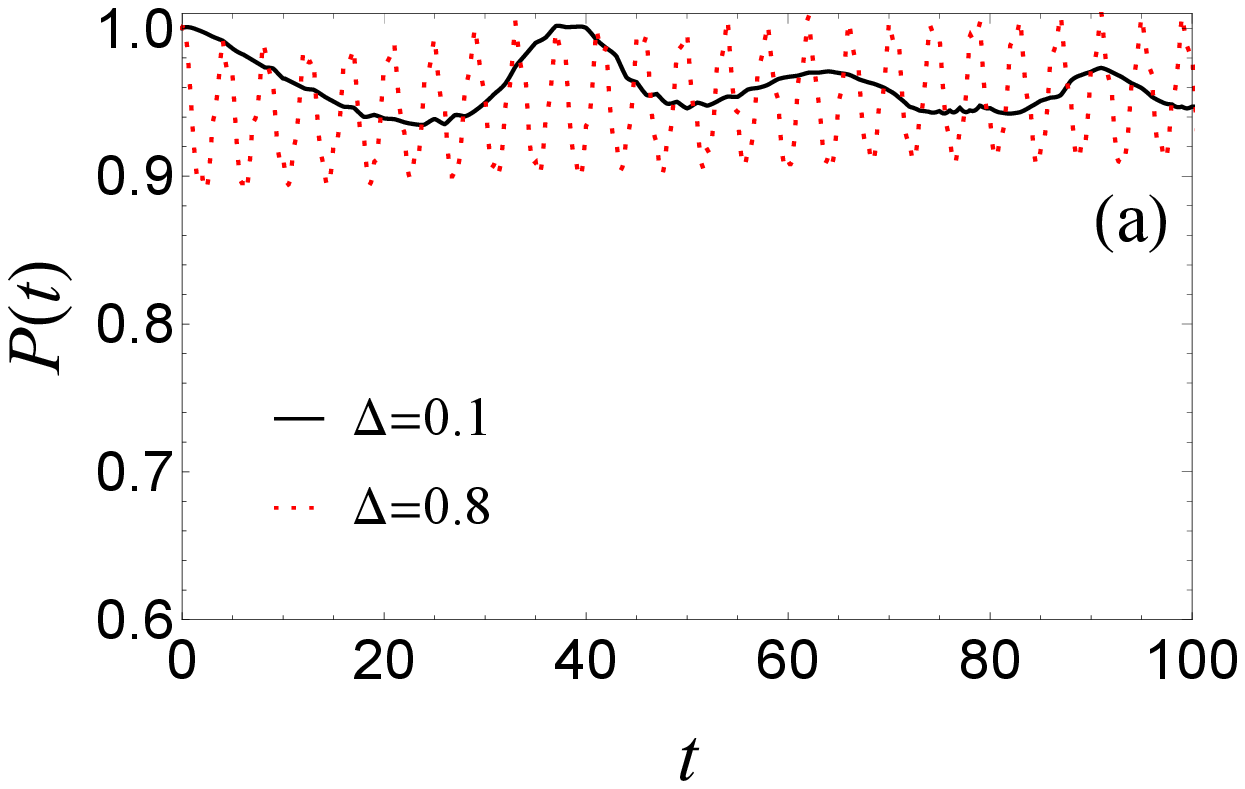}
\includegraphics[width=0.40\textwidth]{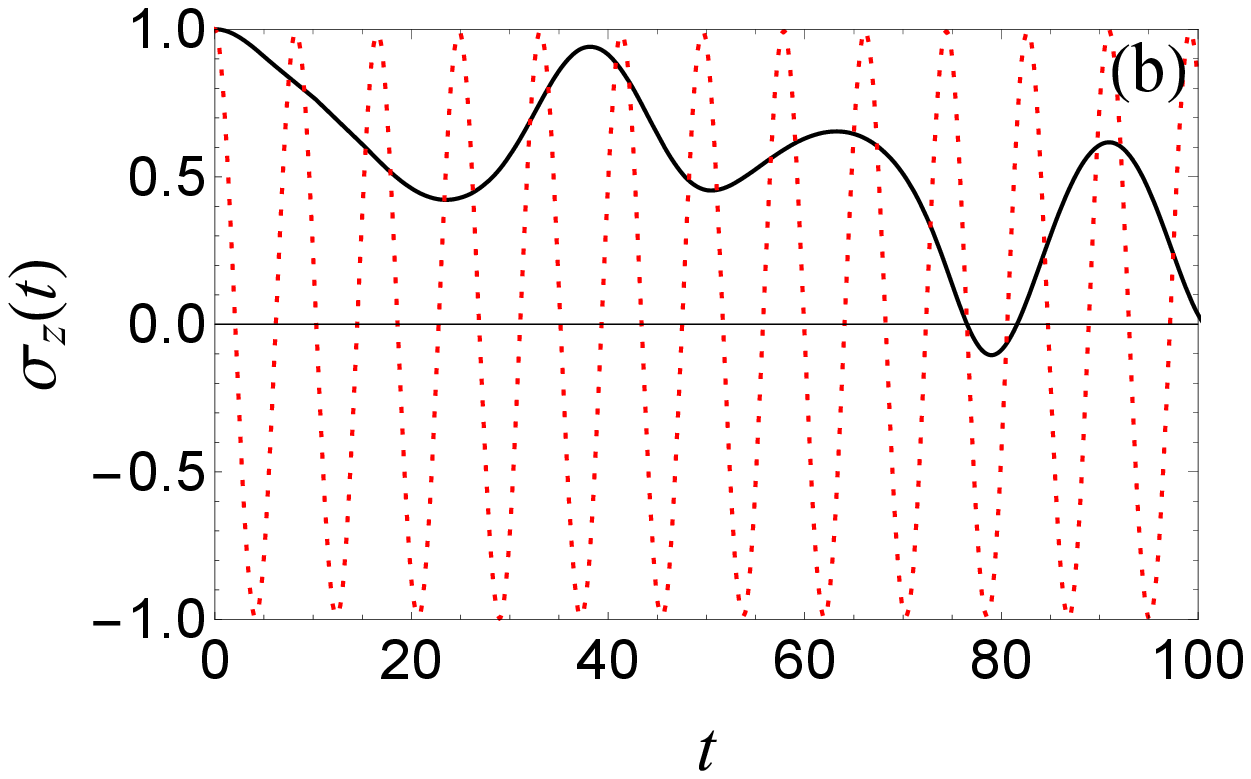}
\end{center}
\caption{
(a) Rescaled purity of the spin state for different $\Delta$ (marked in the plot). The spin is always close to 
the Bloch sphere, $P^{1/2}(t)\approx1$.
(b) Evolution of the {$z$-component of the spin which enters the soliton velocity, see Eq. (\ref{spinvelocity}).
In both plots $\alpha=0.4$ and $g=\wt{g}=-5$.}
}
\label{Fig:1}
\end{figure}

The validity of the adiabatic approximation is corroborated 
by the fact that the spin {state} is always close to a 
pure one (Fig. \ref{Fig:1}(a)), and the spin is close to the Bloch sphere.
In addition, our numerical results show that the shape of the soliton (not shown in the Figures) 
remains practically unchanged up to $t=300.$
As a result, the velocity (\ref{spinvelocity}) self-consistently depends on the evolution of the random 
effective ``magnetic'' field. For a large $\Delta \gg\alpha |k(t)|$, the spin is 
controlled by the Zeeman  
coupling with $\sigma_{z}(t)\approx\cos{(\Delta t)}$ and, according to Eq.  (\ref{spinvelocity}), the velocity behaves 
as $v(t)\approx\alpha \cos (\Delta t).$ For $\Delta \sim {\alpha |k(t)|} $, the 
behavior of observables becomes much more complicated.

Since we consider a narrow soliton with the self-interaction energy conserved under the total spin
rotations, we can introduce a conserved ``low-energy'' quantity  $\epsilon_{0}$ obtained {as the average  (\ref{Hamilton}) 
of the linear part of the Hamiltonian for the adiabatic soliton ${\bm \psi}_{\rm ad}({\bf x})$, i.e.}
  ${\epsilon_0=}k^{2}(t)/2+
\alpha{\sigma}_{z}(t)k(t)+\Delta\sigma_{x}(t)/2+U(X(t)).$
% in Eq. (\ref{expectation}) 
Taking into account that  ${\sigma_{z}(t)}=\cos\theta(t),$ $\sigma_{x}(t)=\sin\theta(t)\cos\phi(t),$
the conservation of $\epsilon_{0}$ {(verified in the adiabatic approximation)} can be presented as 
\begin{equation}
v^{2}(t)-{\alpha^{2}}\sigma_{z}^{2}(t)+\Delta\sigma_{x}(t)=2\left[U({X(0)})-U(X(t))\right],
\label{energy:conserved}
\end{equation}%
with the velocity $v(t)$ given by Eq. (\ref{spinvelocity}). 
The value $U_{\rm inv}$ of the potential that is \textit{sufficient} to invert the soliton dynamics, that 
is $v(t)=0$ at the point where $U(X(t))=U_{\rm inv}$, is 
obtained by minimization of the sum of the spin-related terms in Eq. (\ref{energy:conserved}).
Since this minimum is achieved at $\cos\phi=-1$ and $\sin\theta=\min(\Delta/2\alpha^{2},1)$
we obtain $U_{\rm inv}=U(X(0))+\left(\alpha^{2}+\Delta^{2}/4\alpha^{2}\right)/2$ for $\Delta<2\alpha^{2}$, and 
$U_{\rm inv}=U(X(0))+\Delta/2$ for $\Delta\ge\,2\alpha^{2}.$ 
Nevertheless, these conditions are not \textit{necessary}, and the soliton can be stopped already 
at $U(X(t))<U_{\rm inv}$. At $\Delta=0$ we obtain $U_{\rm inv}=\alpha^{2}/2+U(X(0)),$
corresponding to Fig. \ref{Fig:x_local}(a), where $U_{\rm inv}-U(X(0))=0.083$ for $\alpha^{2}/2=0.08.$ 
\begin{figure}[tb]
\begin{center}
\includegraphics*[width=0.40\textwidth]{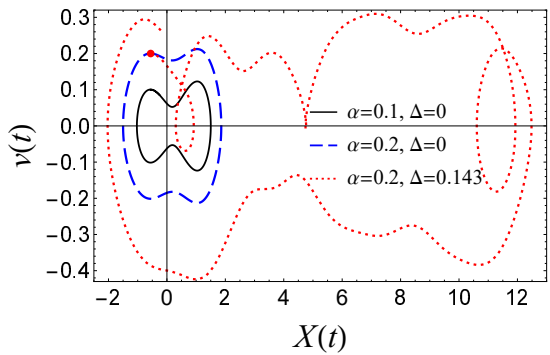}
\end{center}
\caption{Phase trajectory for the values of SOC and
Zeeman field marked in the plot for $t<190$. The filled circle 
describes $\left(X(0),v(0)\right)-$point. The oscillation
frequency for $\alpha =0.1$ is $\omega_{\mathrm{s}}=0.094,$ in a good agreement with 
$\sim{\bar{n}}^{1/4}U_{0}^{1/2}$ estimate.
All plots correspond to $g=\widetilde{g}=-5$.}
\label{Fig:delocalization}
\end{figure}

Before discussing other effects, 
we consider the possibility of experimental realization of the presented system. 
We remind that the physical units of length and time here
are $\xi$ and $t_{\xi}\equiv M\xi^{2}/\hbar$, respectively. The resulting 
coupling constant $g$ is approximately $2a/\xi\times (\xi/\xi_{\perp})^{2}{\cal N},$ where $\xi_{\perp}$ 
is the transversal confinement length, $a$ is the interatomic scattering 
length (e.g., \cite{Mardonov2015a}), 
and ${\cal N}$ is the total number of atoms in the condensate. 
A typical value of $\xi=3$ $\mu{\rm m}$, with  $M$ being the mass of $^{7}{\rm Li}$ atom \cite{Kevrekidis}, 
corresponds to $t_{\xi}\approx 1$ ms and the unit of 
velocity $\xi/t_{\xi}\approx 0.3$ cm/s, 
meaning that our results imply a relatively weak synthetic spin-orbit coupling.  
The relevant time scale of the studied dynamical phenomena, 
being of the order of $100$ $t_{\xi}$, is, therefore,
within the experimental lifetime of an attractive condensate (see e.g. \cite{Trenkwalder2016}). 
For typical values of $a$ of the order of $-5\times 10^{-8}$ cm 
\cite{Moerdijk1994,Pollack2009} and for a strong  
confinement $(\xi/\xi_{\perp})^{2}\sim 10,$
we obtain that the required $g=-5$ can be achieved at ${\cal N}\sim 1.5\times 10^{3}$ particles 
in the condensate. Under these conditions the mean field approach is still well applicable since ${\cal N}
\times(|a|^{3}/\xi\xi_{\perp}^{2})\sim |g|\times(a/\xi)^{2}\ll 1.$

\subsection{Delocalization induced by spin resonance} 

Here we will show that in a certain regime, depending on the oscillation frequency, the Zeeman
coupling and the SOC, the soliton motion can be characterized by a resonance caused by  
spin rotation in the Zeeman field.
Since this resonance occurs in a nonlinear system, it
cannot greatly increase the oscillation amplitude, but it is sufficient to
delocalize a soliton in the case of interest. Although in a disordered
potential the spin resonance is hardly exactly predictable, it is possible
to analyze its effects semiquantitatively. The oscillation 
frequency $\omega_{\rm s}=2\pi/T$ with the 
period 
\begin{equation}
T=2\int_{a}^{b}\frac{dx}{v(x)}, \quad v(x)\equiv\sqrt{2(U(0)-U(x))+\alpha^{2}},
\label{period}
\end{equation}%
where at the turning points $v(a)=v(b)=0.$  
Introducing the critical value $\alpha_{\rm c}^{2}=2[U(x_{r})-U(X(0))],$ we obtain 
$\omega_{\rm s}\sim\omega_{0}/\ln\left[\alpha_{\rm c}\left(\alpha_{\rm c}-\alpha\right)^{-1}\right],$ 
where $\omega_{0}\sim{\bar{n}}^{1/4}U_{0}^{1/2}$ is the 
oscillation frequency near the minimum. Here $x_{r}$ is the position of a strong peak {preventing} the escape of the 
soliton (e.g., in Fig. \ref{Fig:disorder10}, $x_{r}\approx 2.7$).
The resonance between the spin and the orbital motion is expected at $\omega_{\rm s}\sim\Delta.$ 

\begin{figure}[tb]
\begin{center}
\hspace{-0.5cm}\includegraphics*[width=0.36\textwidth]{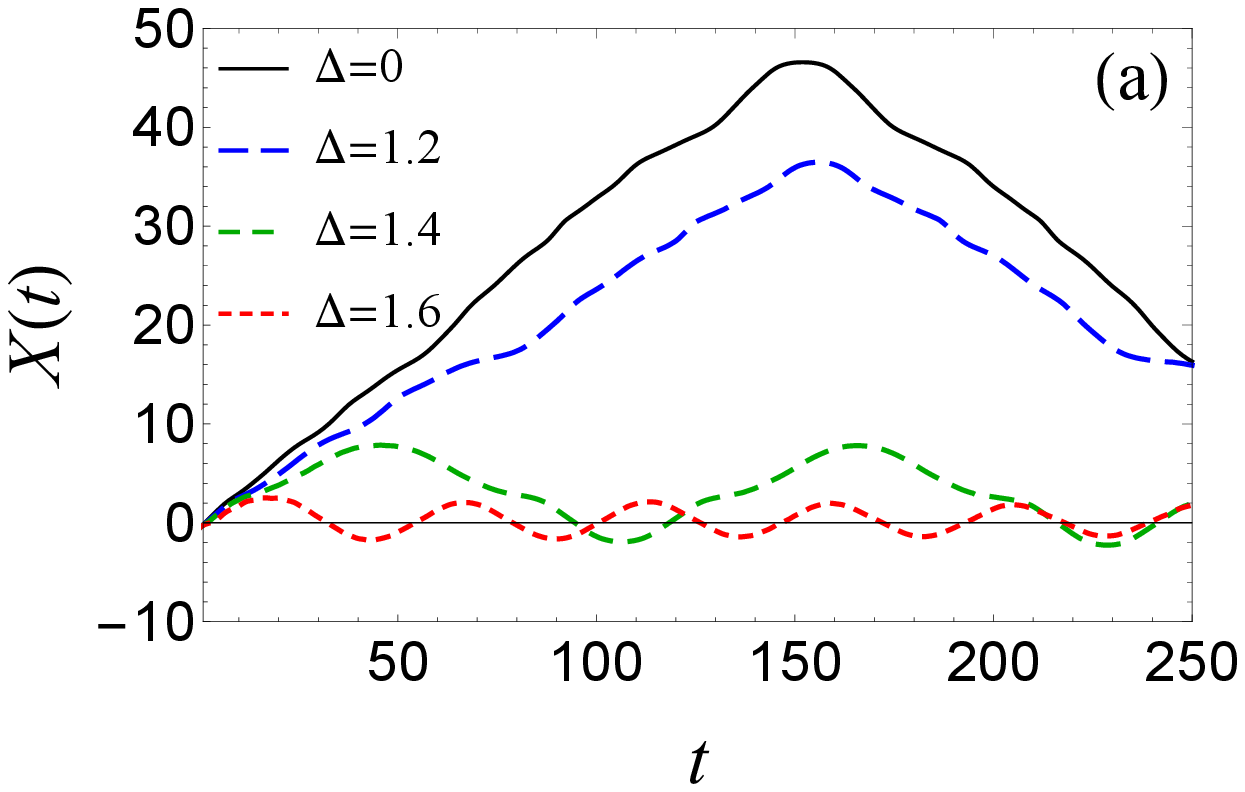} %
\includegraphics*[width=0.40\textwidth]{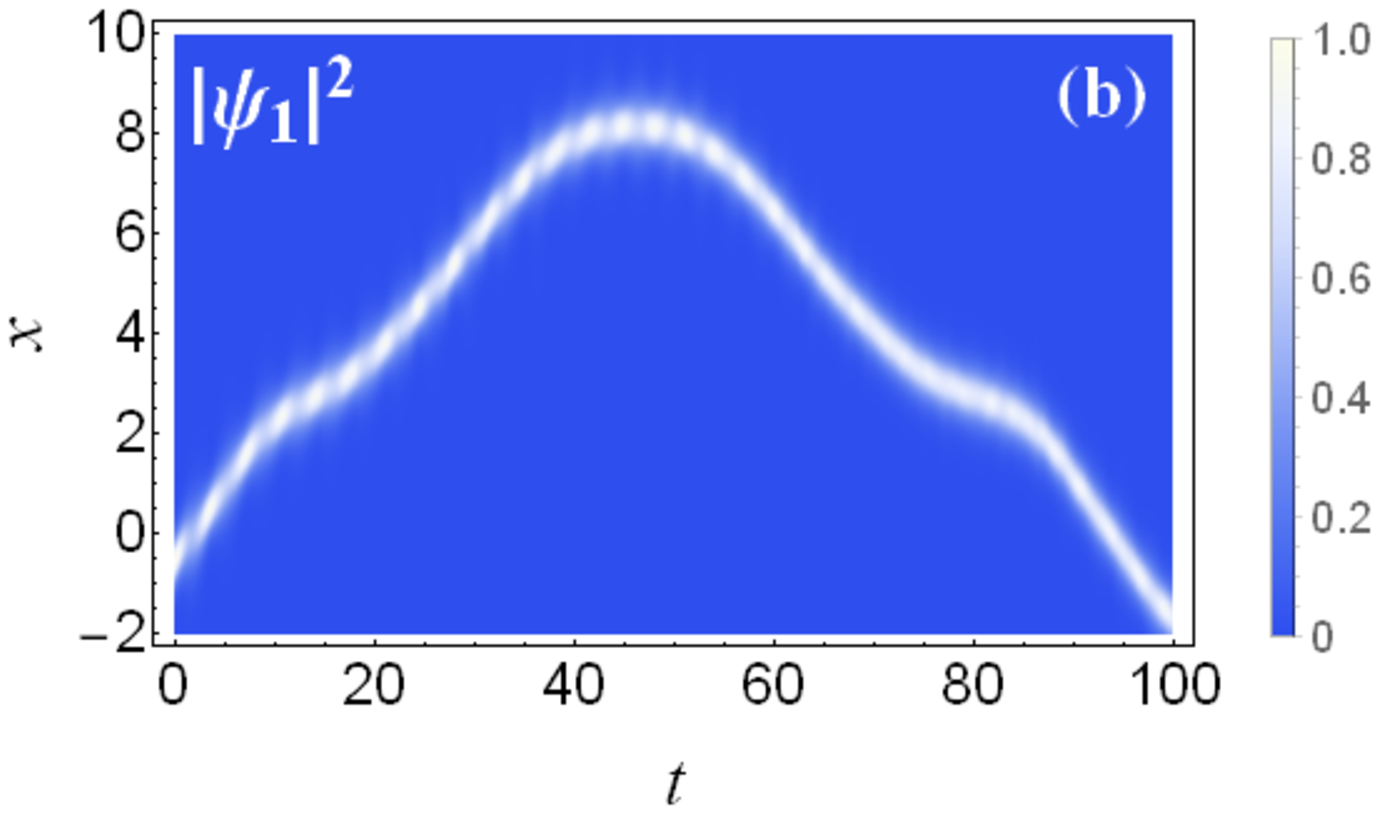} %
\includegraphics*[width=0.40\textwidth]{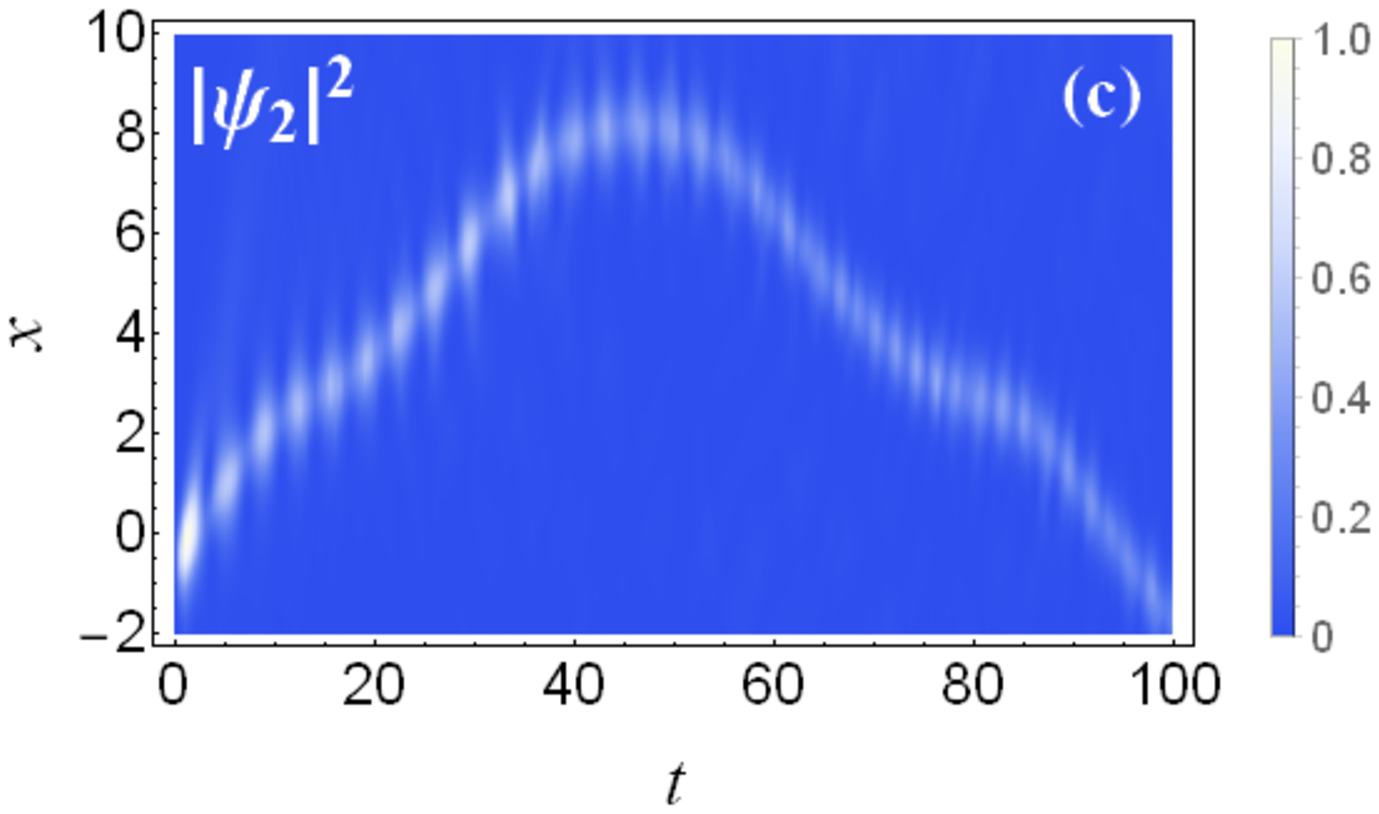} %
\end{center}
\caption{(a) The center of mass position as a function of time for different $\Delta$ and $\widetilde{g}=0,$ $g=-5$.
(b), (c) Density plots of the spinor components at $\Delta=1.4$ in the $(t,x)-$ plane.  
We note two major differences between this figure and Fig. \ref{Fig:x_local} arising due to
different symmetries of the self-interaction. First, here one needs an order of magnitude larger Zeeman $\Delta$ 
to considerably modify the $X(t)-$ dependence.   Second, the density distributions of $\left|\psi_{1,2}(x,t)\right|^{2}$
are broader here with the pattern of $\left|\psi_{2}(x,t)\right|^{2}$ demonstrating a clear stripe-like structure. 
} 
\label{Fig:new_int1_local}
\end{figure}

Figure \ref{Fig:delocalization}, where we show the velocity as a
function of the center of mass position, demonstrates that the interplay 
between the SOC and the Zeeman field can indeed delocalize the soliton. 
For example, for $\alpha=0.2$, the delocalization takes place in the range $0.05<\Delta<0.28$.
Notice that the escape time, velocity, and the direction are random. 

\subsection{Vanishing cross-spin coupling, $\widetilde{g}=0$}

Now we consider the major differences 
brought about by the coupling, {when only the} diagonal self-interaction $g$ is present.
For a direct comparison, we begin with the same initial condition 
as at $\widetilde{g}=g$, that is ${\bm\psi}_{\rm in} (x)$. For this realization of nonlinearity, 
a spin rotation, making $\sigma_{z}(t)$ considerably different from 1,
requires a larger $\Delta$ to provide the energy for population transfer between the 
spinor components $\psi_{1}({\mathbf x})$ and $\psi_{2}({\mathbf x})$. In order to estimate this Zeeman field and to understand 
the interplay between the spin rotation and self-interaction, we assume for the moment $U_{0}=\alpha=0$, and 
consider a ``rigid rotation'' of the wave
function {around the $x-$axis, namely:} ${\bm\psi} (\mathbf{x})=\psi_{0}(x)\left[\cos(\Delta t/2),i\sin(\Delta t/2)\right]^{\rm T}$. 
For this state, the self-interaction energy corresponding to the $H^{\rm int}$ term in Eq. \eqref{Hamilton} becomes
\begin{equation}
E_{\rm int}(t)-E_{\rm int}(0)=-E_{\rm int}(0)\sin^{2}\left({\Delta t}\right)/2,
\label{int1Energy}
\end{equation}%
with the maximum $\max\left(E_{\rm int}(t)-E_{\rm int}(0)\right)$ value $|E_{\rm int}(0)|/2={g^{2}}/12$ achieved
at $t=\pi/2\Delta.$
On the other hand, the Zeeman energy is $\Delta\sigma_{x}(t)/2.$ 
It follows from the comparison of these energies that the spin reorientation by the Zeeman coupling
can provide sufficient energy for the increase in the self-interaction 
if $\Delta\agt\Delta_{c}\sim 0.1{g}^{2}$. Consequently, $\Delta_{c}$ being of the order of one at $|g|=5$,
considerably exceeds the frequencies $2\pi/T\sim \bar{n}^{1/4}U_{0}^{1/2}$ in Eq. (\ref{period}).
As a result, the Zeeman field causes only the soliton localization, as shown in Fig. \ref{Fig:new_int1_local}(a). 

The initial rotation populates the spin down component $\psi_{2}(\mathbf{x})$, which
is initially vanishing, and therefore not self-interacting. 
As a result of small population at $t\ll\,1/\Delta$, the relatively weak self interaction cannot prevent the spread
of this component, and its broadening begins. At the same time, the decreasing (although never vanishing) 
population of the upper component  becomes
insufficient to keep the initial width,  and it is broadening as well. 
The oscillating broadening of $\left|\psi_{2}(\mathbf{x})\right|^{2}$ driven by the Zeeman field is 
seen as the periodic stripe-like structure in Fig. \ref{Fig:new_int1_local}(c).

\section{Discussion and conclusions} 

We have studied the dynamics of the self-attractive quasi one-dimensional Bose-Einstein condensate, 
forming a bright soliton, in a random
potential in the presence of the spin-orbit- and Zeeman couplings.
We have found that for given spin-orbit coupling, the soliton motion strongly depends 
on the Zeeman splitting and on the self-interaction of the condensate. In particular,
the Zeeman interaction can lead to localization or delocalization of the soliton due to the 
spin-dependent anomalous velocity proportional to the spin-orbit coupling. A sufficiently strong  
Zeeman field can cause localization of the soliton near the random potential minima.
If the Zeeman frequency is close to the typical frequency of the soliton oscillations in
the random potential, this resonance can cause its delocalization. 
In the absence of cross-spin interaction, where the Manakov's symmetry is lifted, 
the effect of delocalization due to the Zeeman-induced spin rotation is suppressed 
since a stronger Zeeman field is required here to produce the 
spin evolution sufficient to modify the center-of-mass motion.

\section{Acknowledgments}

We acknowledge support by the Spanish Ministry of Economy, Industry and Competitiveness (MINECO) 
and the European Regional Development Fund FEDER through Grant No. FIS2015-67161-P (MINECO/FEDER, UE), 
and the Basque Government through Grant No. IT986-16. S. M. was partially supported by the 
Swiss National Foundation SCOPES project IZ74Z0\_160527.
We are grateful to Y.V. Kartashov for valuable comments.

\end{document}